\documentclass[aps,prl,preprint,superscriptaddress,showpacs]{revtex4}
\usepackage{graphicx}
\begin{document}
\draft
\def\ee{\varepsilon}
\title{Temperature dependent photon emission spectra of free excitons in phonon field of GaN}
\author{
S. J. Xu }
\email{sjxu@hkucc.hku.hk}
\affiliation{Department of Physics and HKU-CAS Joint Laboratory on New
  Materials, The University of Hong Kong, Pokfulam Road, Hong Kong,
  China}
  \author{G. Q. Li}
\affiliation{Department of Physics and HKU-CAS Joint Laboratory on New
  Materials, The University of Hong Kong, Pokfulam Road, Hong Kong,
  China}
  \author{S.-J. Xiong}
\affiliation{Department of Physics and HKU-CAS Joint Laboratory on New
  Materials, The University of Hong Kong, Pokfulam Road, Hong Kong,
  China}
\affiliation{National Laboratory of Solid State Microstructures and Department of Physics, Nanjing University,
Nanjing 210093, China}

\author{C. M. Che}
\affiliation{Department of Chemistry and HKU-CAS Joint Laboratory on New
  Materials, The University of Hong Kong, Pokfulam Road, Hong Kong,
  China}


\begin{abstract}
Temperature-dependent radiative recombination of free excitons involving one or two LO phonons in GaN is investigated in detail. It is found that both phonon sidebands possess asymmetric lineshape and their energy spacings from the zero phonon line strongly deviate from the characteristic energy of LO phonons as the temperature increases. Furthermore, the deviation rates of one and two phonon sidebands are significantly different. Segall-Mahan theory, taking the exciton-photon and exciton-phonon interactions into account, is employed to calculate the sidebands of one or two LO phonons for free excitons in a wide temperature range. Excellent agreement between theory and experiment is achieved by using only one adjustable parameter (effective mass of free excitons). The obtained effective mass indicates that the free excitons in GaN are likely much lighter than estimated by available theoretical studies.
\end{abstract}
\pacs{78.55.Cr,71.35.-y,71.38.-k} \maketitle

Photon-exciton-phonon interactions are a fundamental many-body issue in solid-state materials, in particular, in direct wide gap semiconductors \cite{toyozawa,rashba}. Such interactions have been recognized to play an essential role in the intrinsic optical properties of these materials for a long time \cite{hopfield,segall,tait}. The studies have provided principal understanding of the photon-exciton-phonon coupling and its effect on absorption and reflectance of photon in nonmetallic materials. On the other hand, for photon emission, especially for phonon-assisted emission, our understanding is considerably insufficient \cite{toyozawa,permogorov}. A main reason for this situation is that such a photon emission or luminescence process involves complicated interpaly of different kinds of quasiparticles. Based on Green's function method, Segall and Mahan \cite{segall1} quantum-mechanically treated the LO phonon-assisted luminescence of free excitons in semiconducting compounds with moderate to reasonably large band gaps. Under the weak coupling approximation, they derived general expressions for one- and two-LO-phonon-assisted luminescence spectra of free excitons. They also attempted to directly compare their theoretical results with available experimental luminescence spectra in CdS and ZnO. In recent years, GaN, which was earlier known to be a wide band gap semiconductor with wurtzite structure \cite{dingle,monemar}, has been extensively investigated due to the first demonstration of the GaN-based bright light-emitting diode with short wavelength by Nakamura in 1993 \cite{nakamura}. However, a current status in the study of GaN is that the empirical technological development is advancing rapidly while a lot of fundamental issues are still far from complete understanding \cite{Johnson}. This is particularly the case for the LO phonon-assisted photon emission of free excitons in GaN. The present growth technology of high-quality GaN epilayers and even free-standing bulks has enabled ones go back to study this issue in detail \cite{kovalev,buyanova,liu,reynolds,leroux,xu,zhang}. Although these experimental studies have significantly enriched our knowledge about the photon-exciton-phonon interactions in GaN, a firm understanding of the LO-phonon-assisted photon emission of free excitons is still lacking. Obviously, such a understanding is of considerable fundamental and technological importance.

In this Letter, we experimentally and theoretically study the temperature-dependent LO-phonon-assisted luminescence of free excitons in GaN. A direct comparison bewteen theory and experiment is made. All the aspects of both one- and two-LO-phonon-assisted luminescence spectra, such as linewidth, asymmetry, and peak shift are well interpreted in a broad temperature range.

The high quality GaN sample used in the present study was a
2.88-$\mu$m GaN epilayer grown on sapphire, followed by a 40-nm GaN nucleation
layer. The sample was prepared with metalorganic vapor phase epitaxy. In the photoluminescence measurements, the samples were mounted on the cold finger of a
Janis closed-cycle cryostat with varying temperature from 3.5 K
to 300 K, and excited by the 325-nm line of a Kimmon He-Cd cw laser
with output power of 40 mW. The emission signal was dispersed by a
SPEX 750M monochromator and detected with a Hamamatsu R928
photomultiplier. The details of the experimental apparatus have been previously described elsewhere \cite{xu}.

As reported in literature \cite{kovalev,leroux}, various excitonic resonant transitions (sometimes called zero-LO phonon lines) keeps dominant in the PL spectra of GaN  even at room tempearture. For the sample studied in the work, 80 K is a dividing temperature below which the resonant line of the bound excitons dominates. When the temperature is higher than 80 K the resonant line of free excitons becomes dominant \cite{xu}. We are primarily concerned with the LO-phonon-assisted radiative decay of the free excitons. In such radiative recombination of free excitons, the generation of photons accompanies with the emission of one or more LO phonons, leading to the so-called LO phonon sidebands. For GaN with wurtzite structure, the characteristic energy of its A$_1$-LO vibration mode is as large as 91 meV. The LO phonon sidebands are thus well resolved from the zero-phonon lines. Due to different coupling strengths of the bound excitons and free excitons with LO phonons, the dividing temperature for the phonon sidebands is only about 25 K \cite{xu}, i.e., the LO phonon sidebands of the free excitons are stronger than those of the bound excitons when the temperature is beyond 25 K. Figure 1 shows two representative LO-phonon-assisted luminescence spectra at 60 K and 100 K. In the concerned spectral range, the one (denoted by 1LO) and two phonon sidebands (2LO) of the free
exciton transition can be clearly resolved. The solid squares represent the measured spectra while the red lines are the theoretical spectra obtained from the use of Segall-Mahan's model \cite{segall1}. It is noted that both peaks are asymmetric. Moreover, the asymmetry of the peaks increases as the temperature increases. Actually, as shown later, the increase rate of asymmetry with temperature are apparently different between two peaks. We also found that the one phonon peak is always broader than that of the two phonon peak in the moderately high temperature range where the two peaks are well defined. More interesting is that the energetic positions of the two peaks strongly blueshift with temperature in different rate. These spectral features unambiguously show that the annihilation of free excitons accompanied with generation of one phonon essentially differs from that with generation of two phonons. In order to quantitatively interpret all these features, we employ Segall-Mahan theory to calcualte one- and two-LO-phonon-assisted luminescence spectra of free excitons.

\begin{figure}
    \centering
        \includegraphics[width=14cm]{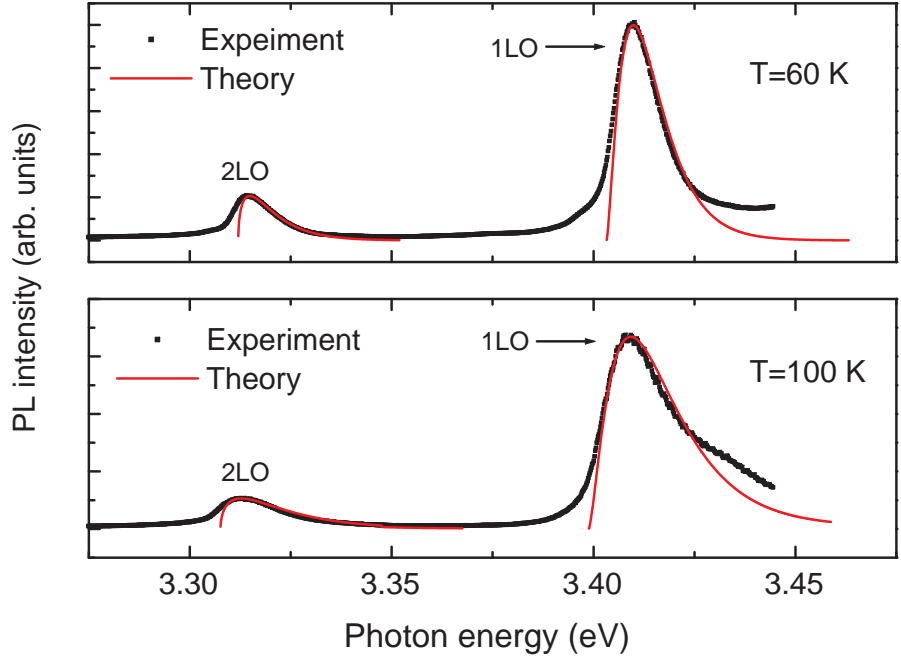}
    \caption{ Two representative LO-phonon-assisted luminescence spectra at 60 K and 100 K. }
    \label{fig:PRLFig1}
\end{figure}

The Hamiltonian for the photon-exciton-phonon coupling system can be written as \cite{segall1}
\[
H=\sum_{\lambda K}\epsilon_{\lambda K}c^{\dag}_{\lambda K}c_{\lambda K}+\sum_{q}\omega_{q}a^{\dag }_{q}a_{q}+\sum_{K}\omega_{K}\alpha^{\dag }_{K}\alpha_{K}
+\sum_{\lambda K}\frac{M_{\lambda}(K)}{\sqrt{\omega_{K}}}\left[c^{\dag}_{\lambda K}\alpha_{K}+\alpha^{\dag}_{K}c_{\lambda K}\right]
\]
\begin{equation}
 +\sum_{\lambda,\lambda^{'};K,q}V_{\lambda\lambda^{'}}(q)c^{\dag}_{\lambda,K+q}c_{\lambda^{'} K}\left(a_{q}+a^{\dag}_{q}\right),
\end{equation}
where $c^{\dag}_{\lambda K}$, $a^{\dag}_{q}$, and $\alpha^{\dag}_{K}$ are the creation operators for exciton, phonon, and photon, $\epsilon_{\lambda K}$, $\omega_{q}$, and $\omega_{K}$ are energies of exciton, phonon, and photon, respectively, $M_{\lambda}$ is the exciton-photon matrix element, and $V_{\lambda\lambda^{'}}$ the coupling strength between exciton and phonon. The optical absorption coefficient can be evaluated from the change rate of photon number operator $N_{K}=\alpha^{\dag}_{K}\alpha_{K}$,
    \begin{equation}
    \label{ee1}
    \partial N_{K}/\partial t=i\left[H,N_{K}\right]=i\left[H_{ex-R},N_{K}\right].
\end{equation}
The general expression of the absorption coefficient for the phtoton-exciton-phonon coupling system can be derived by using the Green's function approach \cite{segall1}
    \[
    \omega W_{abs}(\omega)=\sum_{\lambda}\left|M_{\lambda}\right|^{2}A_{\lambda}
    (\omega)+2\sum_{\lambda\neq\lambda^{'}}M_{\lambda}M_{\lambda^{'}} \left\{\text{Re}\Sigma_{\lambda\lambda^{'}}\left(\omega\right)A_{\lambda}(\omega)\text{Re}G_{\lambda^{'}}(\omega)-\text{Im}\Sigma_{\lambda\lambda^{'}}(\omega)\right.\]
\begin{equation}
\left.\times\left[\text{Re}G_{\lambda}(\omega)\text{Re}G_{\lambda^{'}}(\omega)-\frac{1}{4}A_{\lambda}(\omega)A_{\lambda^{'}}(\omega)\right]\right\},
\end{equation}
where $G_{\lambda}(\omega)$ is the Fourier transformation of the retarded Green's function defined as
\[
  G_{\lambda \lambda'} (K, t-t')= -i \theta (t-t') \langle [c_{\lambda K}(t), c_{\lambda' , K}^{\dag} (t') ] \rangle ,
  \]
($G_{\lambda}=G_{\lambda,\lambda}$), $A_{\lambda}(\omega)$ is the spectral density
\[
  A_{\lambda}( \omega) = -2 \text{Im} G_{\lambda} (0,\omega),
  \]
and $\Sigma_{\lambda, \lambda'} (\omega) \equiv \Sigma_{\lambda,\lambda'} (K=0, \omega) $ is the self energy. Here we set $K=0$ for optical absorption. The transition probability for emission are related to the absorption by \cite{segall1}
\begin{equation}
W_{\text{em}}(\omega) \propto e^{-\omega/k_BT} W_{\text{abs}}(\omega).
\end{equation}

From above expressions the absorption coefficient for the process involving one phonon can be calculated by \cite{segall,segall1}
\begin{equation}
  \alpha_{1} (\omega) = \frac{4\pi e^2 \beta_{A,1} \omega_l (\epsilon_{\infty}^{-1}- \epsilon_{s}^{-1})}
  {4a \hbar c B \sqrt{\epsilon'}} \left( \frac{E_{A,1}}{E_{A,1}-\omega} \right)^2 \left[ \frac{M_{\parallel} B}{\mu_{\bot}\Delta} \right]^{1/2} N(\omega_l) I(\omega),
  \end{equation}
where
 \[
  N(\omega_l) =\frac{1}{e^{\omega_l /k_BT}-1},
  \]
  \[
    I (\omega) =\int_{0}^{-1} dx \left[ 1+\frac{M_{\parallel} - M_{\bot}}{M_{\bot}} x^2 \right]^{-1}
  \]
  \[
  \times \left\{ \frac{2[(1+\eta)^2 -S_e^2]}{S_e [(1+\eta)^2+S_e^2]^2} \text{Im} F_e(\eta) +
    \frac{4(1+\eta)}{[(1+\eta)^2+S_e^2]^2} \text{Re} F_e(\eta) -[e\rightarrow h] \right\}^2.
    \]
Here, $a$ and $B$ are the Bohr radius and binding energy of excitons, $\epsilon_{\infty}$ and $\epsilon_s$ are dielectric constants at high and low frequencies, respectively, $4\pi \beta_{A,1}$ denotes the contribution of the $n=1$ $A$-exciton state to the polarizability, $E_{A,1}$ is its energy, $M_{\parallel}$ is the total masses of exciton in the $c$-axis, $\mu_{\bot}$ is the reduced mass in the direction perpendicular to the $c$-axis, $\epsilon'$ is the dielectric constant at $E_{A1}$ in absence of $n=1$ exciton, $\eta^2 = (E_G-\omega)/B$, $S_{e(h)}= (m_{h(e)\bot}/M_{\bot}) {\bf Q}^{(e(h))}a$, with ${\bf Q}^{e(h)}=(q_x,q_y,q_z[m_{h(e)\parallel} M_{\bot}m_{e(h) \bot}/m_{h(e)\bot}M_{\parallel}m_{e(h) \parallel} ]^{1/2} )$, $\Delta = \omega+\omega_l- E_{A,1}$, and
\[
 F_e(\eta) =_2F_1(2,1-\eta^{-1}, 2-\eta^{-1}, [1-\eta^2+S^2_e-2i S_e\eta] / [(1+\eta)^2+S^2_e]),
 \]
 with $_2F_1$ being the hypergeometric function. The absorption coefficient for the process involving two phonons is \cite{segall,segall1}
 \begin{equation}
  \alpha_{2} (\omega) = \frac{ e^2 \epsilon_s \beta_{A,1} \omega_l^2 (\epsilon_{\infty}^{-1}- \epsilon_{s}^{-1})^2}
  {16a \hbar c B^2 \sqrt{\epsilon'}} \left( \frac{E_{A,1}}{E_{A,1}-\omega} \right)^2 \left[ \frac{M}{\mu} \right]^3 \left( \frac{\langle E_{A,n'}\rangle -E_{A,1}}{\langle E_{A,n'}\rangle -\omega} \right)^2N^2(\omega_l) \sum_n I_n (\omega),
  \end{equation}
 where $M=m_e+m_h$, $\mu^{-1}=m^{-1}_e+m^{-1}_h$, $m_{e(h)} =(m_{e(h),\bot}^2 m_{e(h),\parallel})^{1/3}$, $\langle E_{A,n'}\rangle \sim E_{A,1} + 5B$ is an approximate average energy of exciton levels with $n' \geq 2$, and
\[
  I_n(\omega) = \int_0^{\infty}\frac{dz}{z}\int_{|\beta_n-z|}^{\beta_n+z} \frac{dz'}{z'} |\langle 1 |U(2za^{-1})|1 \rangle \langle 1|U(2z' a^{-1}) |n\rangle
 [(b+z^2)^{-1}-(d+z^2)^{-1} ]
  \]
  \[
  +(d+z^2)^{-1} \langle 1 | U(2za^{-1}) U(2z'a^{-1})|n \rangle + (z\rightarrow z')|^2,
  \]
  with $b=( E_{A,1}-\omega_l -\omega )M/4\mu B$, $d=( \langle E_{A,n''}-\omega_l -\omega )M/4\mu B$, $\beta_n = x_n a/2$, $x_n^2 =2M\hbar^{-2} (\omega +2 \omega_l -E_{A,n})$, and $E_{A,n''}$ being taken as infinity. Here, the matrix element is
  \[
  \langle n|U|1 \rangle =\{ S_1(p_h)+n^{-2}[\frac{1}{4} S_3(p_h) -\frac{1}{12} S_4 (p_h) ]\} -\{p_h \rightarrow p_e\},
  \]
  where
  \[
  S_{\kappa } (p) =-p^{-1} 2^{\kappa +1} (1+p^2)^{-\kappa -1} \exp [-2/(1+p^2)] \{ \sin [2p/(1+p^2)] \text{Re} (1-ip)^{\kappa +1}
  \]
  \[
  +\cos [2p/(1+p^2)] \text{Im} (1-ip)^{\kappa+1} \},
  \]
   with $p_{e(h)} = qam_{e(h)} /M$.

The calculated spectra shown in Fig. 1 are in good agreement with the measured spectra. It can be seen that the position, the line shape, and the assymetry of the sidebands obtained from the theory are in good agreement with the data of the experiment. This implies that the present theory is suitable for the calculation of optical properties of GaN. This coincidence between the theory and the experiment is furthermore illustrated by the temperature dependences of the linewidths of sidebands with one or two phonons extracted from the data and depicted in Fig. 2. Both curves for phonon sidebands with one and two phonons are in agreement of the experimental data. From the approximate expression for phonon emission probability \cite{gross,segall1}
\[
  P_{1(2)}(\Delta)  \sim \Delta^{\gamma_{1(2)}} e^{-\Delta /k_BT} ,
  \]
  where the subscript 1 or 2 is for the emission of one or two phonons, and $\gamma_1 =3/2$, $\gamma_2=1/2$, the linewidths are approximately $2.958k_BT$ and $1.795k_BT$ for emissions of one and two phonons, respectively. These approximate results are shown in Fig. 2 with light lines. Compared with the approximate results, the present theoretical calculation includes intermediate excitonic states with $n>1$ and gives an evident correction for $T>0$. For the one-phonon process this correction is negative from the sign of the matrix elements, while for the two-phonon process it is positive due to the second-order terms in the perturbation theory.

The temperature dependences of the energy distances of the sidebands with one and two phonons from the zero-phonon line obtained from the theory and from the experiment are shown in Fig. 3. An excellent agreement between the theory and the experiment is also achieved. The energy distances are decreased with the increasing temperature, because the population of polaritons is shifted toward the high-energy states by increasing the temperature. From the approximate theory the energy distance is always the energy of phonons involved. Our theoretical calculation gives negative corrections to the approximate values for both the one-phonon and two-phonon processes. These corrections are originated from the renormalization of the energy levels due to the scattering to and from the intermediate excitonic states. It can be seen that the correction to the one-phonon process is larger than that of the two-phonon process, implying the perturbative nature of the scattering in which the correction is decaying with the order. For both cases the correction increases with increasing the temperature, because at higher temperatures more intermediate excitonic states are involved. 

\begin{figure}
    \centering
        \includegraphics[width=12cm]{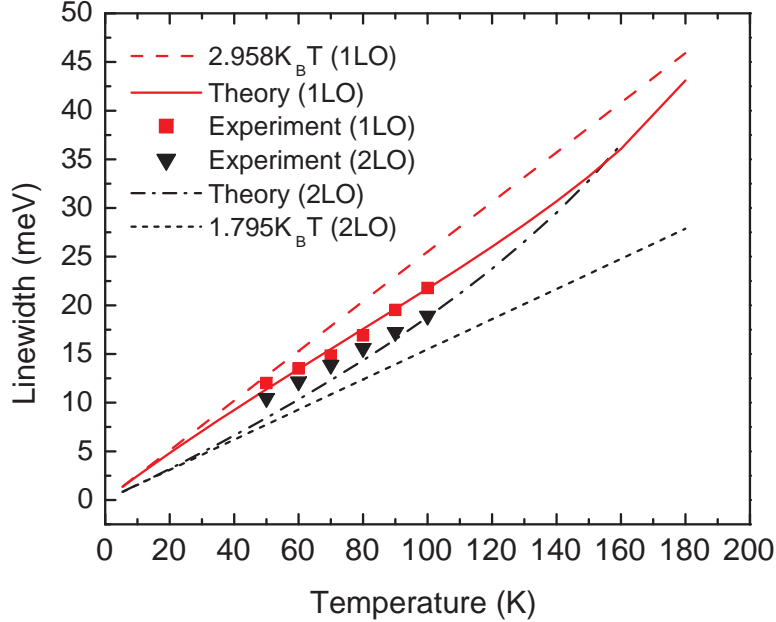}
    \caption{Linewidth of sidebands with one and two phonons as functions of temperature. The lines are the calculated results, and the filled symbols are the measured data. }
    \label{fig:PRLFig2}
\end{figure}

\begin{figure}
    \centering
        \includegraphics[width=14cm]{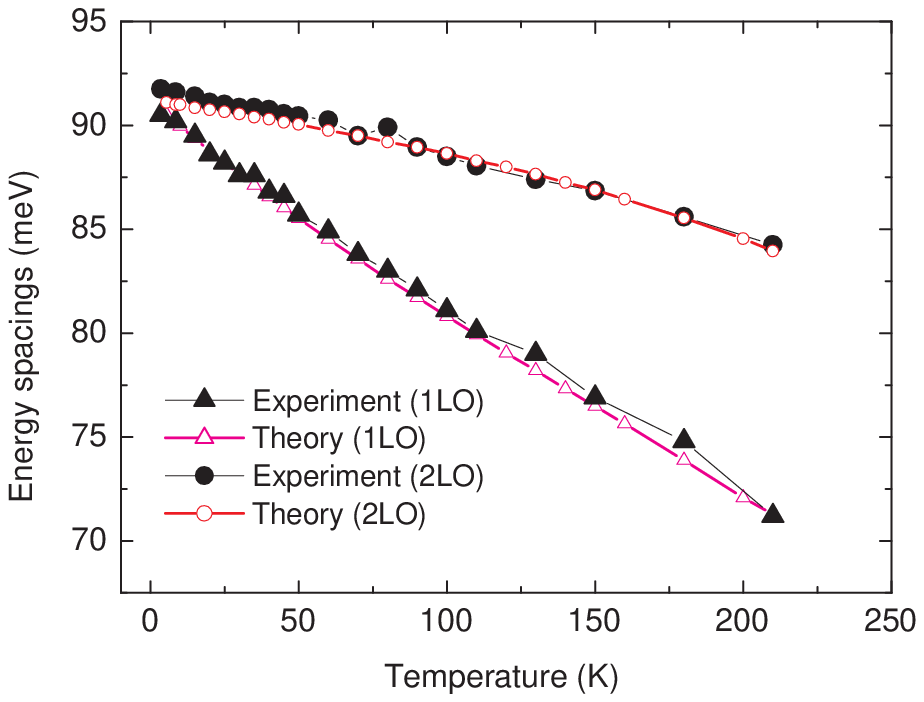}
    \caption{Temperature dependence of the energy spacing from the sidebands with one and two phonons to the zero-phonon line obtained from the theory and from the experiment. }
    \label{fig:PRLFig3}
\end{figure}

In the calculations we use the following parameters for GaN \cite{nakamura,xu,bougrov,yeo}: $m_e=0.2m_0$, $m_{h\bot}=0.5m_0$, $m_{h\parallel}=0.6m_0$, $B=0.0254$eV, $\omega_l =0.0912$eV, $\epsilon_s =8.9$, $\epsilon_{\infty} = 5.35$, $E_{A,1}(T=60K)=3.4944$eV, $E_{A,1}(T=100K)=3.4900$eV, and $4\pi \beta_{A,1} = 0.0066$. It should be emphasized that all the parameters, except the effective hole mass, are taken from those of the material. The best agreement is achieved between the theory and the experiment when we choose $m_{h\bot}=0.5m_0$, $m_{h\parallel}=0.6m_0$, which are much lower than the results of band calculations, $m_{h\bot}\sim m_{h\parallel}\sim 1.5m_0$. From the excellent agreement between the theory and the experiment, we suggest that the values used here are the better ones. This is supported from the experimental value of the binding energy. The relation between the binding energy and the reduced mass of an exciton is \cite{pyu}
\[
      B= \left( \frac{\mu}{m_0 \epsilon_s^2}\right)\text{Ry} ,
      \]
with Ry=13.6 eV. From this relation and measured values of $B$ and $\epsilon_s$ listed above, the reduced mass is $\mu = 0.148m_0$. Since the effective mass of electron has been well measured as $m_e= 0.2m_0$,  from $\mu$ one can determine the effective mass of hole as $m_h \sim 0.57m_0$. This supports the values of $m_{h\bot}$ and $m_{h\parallel}$ used here for giving the best consistence of the theory and the experiment. Thus, the present measurement of the temperature dependent photon emission spectra provides another way of determining the hole effective mass in GaN, and the obtained result coincides the relation between the binding energy and the reduced mass, although it is different from the results of the band calculations. 

As a summary, we experimentally and theoretically investigate the temperature-dependent radiative recombination of free excitons involving one or two LO phonons in GaN. It is found that both phonon sidebands possess asymmetric lineshape and different temperature dependence. The calculated spectra from the Segall-Mahan theory are in excellent agreement with the measured results in a wide temperature range. This study provides a method of determining the effective mass of holes in GaN.

The work in Hong Kong was supported by the HK RGC CERG Grant (No.
HKU 7036/03P). S-JX gratefully acknowledges research support
from Chinese National Foundation of Nature Sciences (Nos. 60276005 and 10474033). 
SJX wishes to thank Dr. W. Liu and Prof. M. F. Li for providing GaN sample.

\end{document}